# Do We Still Need Coins? The Role of Payment System Innovation, Pandemic, and Coin's Purchasing Power, on Coin Demand in Indonesia

Wishnu Badrawani[1*], Elsa Dyahpitaloka[1], Ahmad F.F. Alanshori[1], Imam Mukhlis[2]
[1]*Bank Indonesia, Jakarta, Indonesia*
[2]*Department of Economics, State University of Malang, East Java, Indonesia*

***Abstract***

*Purpose*– This study investigates the relationship between coin demand, payment system innovation, COVID-19, and a coin's purchasing power. The rapid advancement of payment platforms combined with their high adoption during the pandemic, has positioned non-cash payment methods as a possible complement or alternative to coin money for transactions. However, there is notably limited coin money-related research in the economic literature.

*Design/methodology/approach*– The autoregressive distributed lag (ARDL) bounds test methodology's co-integration approach was applied using monthly data from 2011 to 2022.

*Findings*– Our findings reveal a long-term relationship between coin demand and its determinants, namely payment innovations, pandemic, coin depreciation, and income. Despite the swift advancement of payment innovations and the rapid increase of non-cash payments, coins remain vital to the economy and are unlikely to become obsolete in the near future.

*Practical implications*– Policymakers must understand the driving factors of coin demand in both economic and non-economic contexts to improve coin production-related issues and coin circulation policies. Reviewing the denomination structure of the Rupiah is crucial in addressing the problem of ineffective coin circulation in the economy.

*Originality/value*– This is the first study to examine coin money's relationship to payment innovations and the pandemic, particularly in emerging countries like Indonesia.

*) Corresponding author's email address: w_badrawani@bi.go.id
The opinions presented in this paper do not constitute the official position of Bank Indonesia but are the author's private opinions.

## 1. Introduction

Initially, money is considered "neutral" and does not affect the size of the real economy. Since the barter system, the use of money has served the objective of satisfying a double coincidence of demands that facilitated the exchange of goods. Under the assumption of perfect information and negligible transaction costs, its function as a value store is considered very limited. The relevance of money demand in macroeconomic policy formulation has been emphasized by considering Keynes' liquidity preference theory, which asserts that transactional, speculative, and precautionary motives influence money demand. The notion of "money demand" encompasses a range of monetary manifestations, including banknotes and coins representing physical currency, financial assets held in bank accounts, quasi-money, and digital currencies.

The influence of financial innovation on money demand has become a crucial topic in recent times. Dunne & Kasekende (2018) and Mlambo & Msosa (2020) suggest that financial technology advancement has an adverse effect on money demand. Lubis et al. (2019) also suggest that using electronic and card payment methods leads to decreased cash payments, with substitution effect were more prominent for lower denominations (Shy, 2023). Thus, it is expected that the necessity of small denominations will be decreased in the future. Interestingly, coins and overall money in circulation in Indonesia have increased for more than a decade, even during and after the COVID-19 pandemic (Figure 1).

Coins have played a significant role in Indonesia's monetary and economic system for centuries. Ancient coins with various materials and denominations had already circulated in the economy prior to the country's independence. The primary function of these coins was to enhance local trade within the archipelago and facilitate international trade. The participation of Indonesia in ancient global commerce networks notably impacted the assortment of coins employed inside its monetary system (Munandar, 2020).

The relationship between the policy rate and coin demand from 2011 to 2022 is shown in Figure 1(a). Figure 1 (b) illustrates the usage of ATM and debit cards, and the overall transactions made through cards and digital payments. These transactions have shown a consistent rising trend since 2011, with seasonality and a decline during the pandemic. Electronic money experienced a significant increase during and after the pandemic crisis.

---

Insert Figure 1

**(a) Coin Money in Circulation and Policy Rate, (b) Non-Cash Payment Instruments**

---

Figure 1 shows that the use of cash remained high despite the rapid increase in non-cash payments and the containment measures in response to the pandemic, which included social distancing, temporary closures of workplaces and schools, social lockdowns, and restrictions on the use of cash to prevent virus transmission through contact with ill individuals. Cash in circulation has continued to rise for over a decade, notwithstanding the pandemic and the swift advancements in non-cash payment methods. This phenomenon has been corroborated by studies conducted in various countries (Lalouette & Esselink, 2018), including Australia (Guttmann et al., 2021), Japan (Honda & Uesugi, 2022), the United States, and several other industrialized nations (Rogoff & Scazzero, 2021). The rapid advancement of payment systems has expanded consumer options for selecting payment methods, potentially complementing or substituting coins in transactions. This further emphasises the need to examine the relationship between payment innovation and the demand for Rupiah coins.

The utilisation of coins presents a more significant number of challenges compared to banknotes. Certain regions in the country have witnessed consumer and merchant rejection of the acceptance of coin money (Putri & Nugroho, 2023). These results were consistent with Lahanta & Zulfadhli (2019) and paralleled a phenomenon known as the "burden of holding coins" that occurred in developed nations like Canada (Chen et al., 2021). Despite the extensive examination of the impact of financial innovation on money demand in high-income countries, studies pertaining to coin currency remain few. This is somewhat unexpected, considering the substantial growth in financial technology has influenced people's transactional behaviour, specifically regarding the declining household preference for coin money and the increasing inquiry into the necessity of coins in the digital age. This implies that the discourse surrounding the demand for money should additionally encompass coin money.

We fill this gap by using coin money as a monetary aggregate with various measures of payment system innovation, COVID-19, and coin purchasing power to provide a more comprehensive view of Indonesia's stability of coin money demand. Our research is distinguished from existing studies that mostly examine overall money demand and its determinants. This study aims to verify whether the Indonesian economy still requires coins. Our study contributes to the literature by shifting our focus from general money demand to coin money. This article offers policy recommendations for authorities, suggesting improvements in projecting coin demand and circulating coins within the economy.

The remainder of the paper is structured as follows: Section 2 provides an overview of the relevant literature. Section 3 covers the data, model design, and estimation. Section 4 presents and discusses the empirical results. Finally, Section 5 discusses the conclusions and recommendations.

## 2. Literature Review

### *2.1 Theoretical review*

Various economic theories have explained the role of money demand within the real economy. Classical economic theory explains money demand by examining the relationship between price levels and transaction values in the economy, thereby introducing the concept of money velocity (Fisher, 1911). The Fisher money equation asserts that, ceteris paribus, an increase in the money supply results in a corresponding rise in the price level, leading to a decline in the value of money, and conversely. Money is thus regarded as neutral, serving to facilitate transactions.

John Maynard Keynes (1936) asserts that the demand for money is affected by speculative motives related to interest rate (r), as well as transaction and precautionary motives that depend on the level of income and wealth. Many subsequent theories were influenced by this concept, including those developed by Milton Friedman, who stated that the opportunity cost of holding currency is equal to the potential profit from investing in interest-bearing assets, stocks, bonds, and other investments (Poff, 2023). The demand for money declines when the marginal rate of substitution between money and alternative assets diminishes; consequently, an increase in the return on other assets will lead to a reduction in the money demand. The basic money demand function commonly is expressed as follows:

$$M_t/P_t = L(y_t, R_t) \quad (1)$$

Where $M_t/P_t$ is the real money demand of the household during period $t$, $y$ is the household's consumption during period $t$, and $R$ represents the interest rate of certain assets. In this regard, Friedman posited that changes in the return on bonds, stocks, and goods would cause corresponding fluctuations in the money supply. Additionally, the impact of monetary policy would be delayed (Dupor, 2023).

### 2.2 Empirical Studies

Extensive empirical studies have been conducted on money demand related research, including the necessity of analysing money demand within the framework of inflation-targeting regimes. Although many countries regarded the inflation targeting framework (ITF) as the leading approach to monetary policy and moved away from monetary targeting, the ECB continues to consider money a significant variable in its monetary policy, and extensive research on money demand is still being conducted (Greenwood, 2022).

Sun, Yin, and Zeng (2023) assert that wealth, acting as a financial transaction motive, and the annual change in the unemployment rate, serving as a precautionary motive, positively impact the demand for money. Dritsaki and Dritsaki (2020) discovered evidence of both a short- and long-term relationship between the factors of money demand, namely income, interest rate, and inflation. These results support Bissoondeeal, Binner, and Karoglou (2023) with the inclusion of economic uncertainty and stock market volatility using the Euro area, the United States, and the United Kingdom data. Dunne & Kasekende (2018) found a cointegration relationship between money and real output in African countries, both narrow and broad money.

Previous studies on money demand in Indonesia used a variety of factors and approaches, indicating a cointegration relationship and similar findings, for example, Wicaksono et al. (2024) and Lubis et al. (2019). Zams (2018) demonstrated the presence of a structural break that supported the long-run relationship between real money demand and its determinants. The standard money demand function in Indonesia can be written as follows:

$$MD_t = \beta_0 + \beta_1 y_t + \beta_2 r_t + e_t \qquad (2)$$

where $MD_t$ represents real money demand, $y_t$ is real income, and $r_t$ represents the interest rate.

Studies conducted on payment systems and financial innovations have consistently shown that they have a substantial influence on the demand for money in both developed and developing countries. Using the shopping time model, McCallum & Goodfriend (2018) explained that technological advancements in the payment system affect people's behaviour in using cash, which ultimately impacts the demand for money. Lippi & Secchi (2009) found that the adoption of cash withdrawal technology led to a decrease in the interest elasticity of the money demand curve, and advancements in cash withdrawal technologies counterbalanced the expected increase in cash demand caused by decreased interest rates. Evidence suggests that technological advancements have an adverse effect on the demand for money, according to Bashayreh et al. (2019) who utilised panel data from banks in Jordan and 230 countries, respectively. This finding is consistent with Lubis et al. (2019), who suggest that debit card transactions and payment infrastructure negatively impact currency demand.

Mlambo & Msosa (2020) and Guttmann et al. (2021) found that financial innovation, proxied by mobile subscriptions and ATM transactions, has a negative effect on money demand. Chucherd et al. (2019) support these findings by using electronic payment instruments, such as e-wallets, credit cards, debit cards, the Internet, and mobile banking, on money demand. The innovation of payment instruments plays a significant role in promoting financial inclusion and meeting the demand for smaller currency denominations in retail transactions (Alwahidin et al., 2023; Frost et al., 2021).

The study of coin usage by Lahanta & Zulfadhli (2019) found that several factors affect the acceptance or rejection of coins, including the devaluation of coins, low use of cash by individuals, and regulations. This finding corroborates Chen et al. (2021) and Spector (2019) who found that carrying excessive amounts of coins might provide difficulties, particularly when doing larger value transactions. Thus, assessing the currency denomination structure is vital to ensure that money functions efficiently as a means of payment throughout the economy (Manikowski, 2017). Further challenges related to maintaining currency circulation in the economy were the rise of hoarding behavior and a tendency for negative seigniorage due to high production costs.

The COVID-19 substantially impacted cash usage, primarily due to economic activity restrictions imposed as containment measures (Auer et al., 2020; Rowan & Laffey, 2020). These findings corroborate Baldwin & Weder (2020) and Rogoff & Scazzero (2021), highlighting how the ongoing surge in cash circulation contrasted with the swift growth of electronic payments. Guttmann et al. (2021) and Assenmacher et al. (2019) discovered that the increased banknote demand during the pandemic might be attributed to individuals' desire to hoard cash for precautionary or wealth-preservation motives. In Sweden, the amount of currency on hand increased during economic turmoil, similar to the 1990s when the Bretton Woods system collapsed. This finding corroborates Honda & Uesugi (2022) in their examination of corporate cash-holding behaviour. Chen et al. (2021) found that most individuals hold cash in medium to lower-value denominations, and 74% of participants anticipate continuing to use cash in the future.

According to the review above, to the best of our knowledge, limited studies examine the money demand that focus on coin demand wich include the metric of payment innovation and the COVID-19. An analysis of coin demand and its determinants, including payment innovation, constitutes a significant contribution to the literature.

## 3. Data, Model Specification and Estimation

### *3.1 The Data*

The datasets utilised in this study are derived from secondary data, specifically the Indonesian Economic and Financial Statistics (SEKI) and Payment System Statistics (SPIP) provided by Bank Indonesia. These datasets were employed to conduct an empirical analysis to address the research objectives with consideration of data availabilities. We use monthly data from 2011 to 2022. We transformed all variables into natural logarithmic form, except for interest rate and dummy variables, to enhance the accuracy and reliability of the estimates.

### *3.2 Model Specification*

Following Mera et al. (2020) with modifications based on the prior studies, we develop the following model to estimate the demand for coin money as follows:

$$CMD_t = \beta_0 + \beta_1 y_t + \beta_2 r_t + \beta_3 valco_t + \beta_4 dum_t + \beta_5 PSI_t + e_t \qquad (3)$$

In this study, *CMD* refers to coin money demand, while ***y****, r,* and *valco* represent real income (GDP), the policy rate, and the value of the highest Rupiah coin with respect to gold, respectively. The measurement of the payment system innovation *(PSI)* is determined by the usage of ATM/Debit card transactions *(ATMD)*, the total of card-based and digital payment instruments *(APMK)*, and electronic money *(EM)*. There are two dummy variables *($dum_t$)*, which are the impact of the pandemic *(DUM_C19)* and the holiday season during Eid al-Fitr *(DUM_EID)*. These combinations collectively provide a total of three models. A detailed table containing the descriptions of the variables is available in Table A1 of Appendix A.

### *3.3 Cointegration test: ARDL approach*

This study employs an ARDL bounds testing approach to determine if there is a long-term relationship between the variables. The following specifications illustrate the approach we will use for cointegration:

$$\begin{aligned}\Delta CMD_t = {} & \beta_0 + \beta_1 CMD_{t-1} + \beta_2 y_{t-1} + \beta_3 r_{t-1} + \beta_4 valco_{t-1} + \beta_5\, dum_{t-1} \\ & + \beta_6\, psinv_{t-1} + \sum_{i=0}^{N} \theta_i\, \Delta MD_{t-i} + \sum_{j=0}^{N} \theta_j\, \Delta y_{t-j} + \sum_{k=0}^{N} \theta_k\, \Delta r_{t-k} \\ & + \sum_{l=0}^{N} \theta_l\, \Delta valco_{t-l} + \sum_{m=0}^{N} \theta_m\, \Delta dum_{t-m} + \sum_{n=0}^{N} \theta_n\, \Delta PSI_{t-n} + \varepsilon_t \end{aligned} \qquad (4)$$

where $\beta_0$ denotes the drift, $\varepsilon_t$ is the vector of the white noise error process, which is assumed to be serially uncorrelated and homoscedastic. Δ is the first difference operator, N is the ARDL model maximum lag order, and $\theta_l - \theta_n$ and $\beta_0 - \beta_6$ are the long-run and short-run parameters, respectively. All the variables used in Equation (4) are defined as in Equation (3). The long-run relationship can be tested using the F-test statistic obtained from the bounds test. The standard

Wald (or F-test) statistic is employed to assess the joint significance of the one-period lagged level variables in Equation (4) based on the following hypotheses:

$$H_0: \beta_1 = \beta_2 = \beta_3 = \beta_4 = \beta_5 = \beta_6 = 0$$

$$H_1: \beta_1 \neq \beta_2 \neq \beta_3 \neq \beta_4 \neq \beta_5 \neq \beta_6 \neq 0$$

The null hypothesis of no cointegration is rejected when the F-statistic exceeds the upper bound, whereas it is accepted when the F-statistic goes below the lower threshold. However, it is inconclusive if it falls inside the range of critical values, specifically between the lower and upper bounds (MacKinnon, 2010).

### *3.4 Diagnostic Test and Parameter Stability*

In conjunction with the ARDL method, a thorough battery of diagnostic tests was employed to assess the performance of the resulting models, encompassing serial correlation, normality, heteroscedasticity, and functional form, to ascertain the reliability and robustness of the parameter estimations for policy analysis. According to Schweikert (2020), the empirical evidence indicates that the existence of cointegration does not necessarily ensure the stability of the estimated coefficients. Thus, stability tests were conducted using the cumulative sum (CUSUM) and cumulative sum of squares (CUSUMSQ) methods, which rely on the recursive regression residuals.

## 4. Results and Discussions

### *4.1 Descriptive Statistics*

Table 1 displays the results of the descriptive statistics and unit root test using the Augmented Dickey-Fuller (ADF) unit root tests. All variables exhibited stationarity, either at the level or at first differencing, indicating the absence of any I (2) variable. Therefore, it is suggested that an examination employing the ARDL models be undertaken.

Insert Table 1

**Descriptive Statistics**

## *4.2 Long-run and Short-run Relationship*

### *4.2.1 Bounds Test for Cointegration*

The ARDL model employs the bound test for cointegration in the computation of the F-statistic to examine the presence of a long-run relationship between the coin demand and the dependent variables which is shown in equation (3). The results of the bound test for cointegration revealed that the estimated F-statistics for Model 1 to Model 3 tests exceeded the upper critical bound value, as displayed in Table 2.

Insert Table 2

**ARDL Bound Test Results for Cointegration**

Therefore, the null hypothesis of non-cointegration can be rejected in all three models that represent different forms of payment instruments, leading to the conclusion that there is a long-run relationship between coin demand and its determinants.

### *4.2.2 Long-run result*

Table 3 displays the long-run relationship results for the coin demand model across several specifications, as indicated in columns (1), (2), and (3). The results of the long-term estimated model show that several observed variables, including income, interest rate, exchange rate, the presence of the holiday season and COVID-19, and the value of coins, significantly impact the demand for Rupiah coins.

Insert Table 3

**Long-Run ARDL Estimation Results**

In the long-run models, the coefficient indicators for most explanatory variables are consistent with previous empirical studies (Adil et al., 2020; Mlambo & Msosa, 2020; Zhan et al., 2023).

The long-term analysis reveals that the variables representing **payment innovation**, ATM/debit card, and overall retail payment, except for e-money, significantly impact the coin demand in Indonesia. These findings follow previous studies by Chen et al. (2021), Lubis et al. (2019), and Zhan et al. (2023), with debit cards as the most popular non-cash payment mode during and after the pandemic. However, we discovered that the sign for payment instruments was different from most literature. This finding means that non-cash instruments are associated with increased Rupiah coin demand, raising issues about the reasons. We discovered that coin money was not adequately re-circulated throughout the economy according to focus group discussions with officials from the central bank's regional offices and commercial bank officials. This condition refers to the very few returns of coins Rupiah to Bank Indonesia because of limited business entities re-circulating the coin due to handling cost issues, customer hoarding behavior, and limited use of coins for transactions, possibly due to the negligible economic value of coins (Badrawani et al., 2024). Meanwhile, the ongoing need for coins for small transactions among retailers, merchants, and small economies, combined with the decline in coin returns to the central bank (refer to Appendix A2), resulted in an upward trend in overall coin demand, as the statistics indicate. The increased demand for coins is also seen in developed countries (Perkins, 2019), whereas limited coin use is often observed in countries with low coin value (Sahar & Setiartiti, 2016; Veber & Brosch, 2013).

**The COVID-19** dummy variable was statistically significant in affecting the coin demand. This finding aligns with Honda & Uesugi (2022) and Kotkowski (2023), who posit that the surge in cash reserves during the pandemic is attributed to precautionary motives and cash-holding culture. According to Duc et al. (2022), the rise in money supply is attributed to money creation by central banks and commercial banks, supported by the purchase of government assets. Social safety nets and government cash transfers also contribute to this increase (Shaefer et al., 2022).

However, regarding the remarkably little worth of coin money, which appears unsuitable to represent the precautionary motive, there could be a non-economic factor influencing this situation. The long-term analysis demonstrates that **the declining value of the coin** over time (valco), which captures the inflation effect of coin value, is statistically significant at the 5% level for Model 3. This result indicates that a percentage rise in the depreciation of the coin's value reduces the average demand for coin money by 0.28 to 0.53 per cent. These findings support prior research regarding the importance of preserving the value of physical currency, including coins, for practical use in daily transactions; for example, see Manikowski (2017), Arshadi (2019), and Badrawani et al. (2024).

The **income** coefficient exhibits a high degree of significance at the 1% level, as ascertained from the long-term analysis. Furthermore, it exhibits the expected positive sign, indicating that a one-unit increase in income corresponds to a 1.84, 1.85, and 2.47 increase in coin demand for all models. The result is consistent with the literature that suggests that economic agents' demand for currency will increase in response to rising income. The estimated coefficient value is greater than one and positive, slightly higher than previous research on money demand in Indonesia, such as that of Zams (2018) and Lubis et al. (2019), but following Bangura et al. (2022). The disparity in results can be attributed to the distinct nature of our dependent variable, which focuses on coin money rather than narrow or broad money. Additionally, non-economic factors such as hoarding behaviour and the cash-handling practices of financial institutions may have influenced the outcome. This circumstance may introduce a biased interpretation if we solely consider the data on coin demands and GDP, which may not accurately represent the demand for Rupiah coins within the economy.

The findings indicate that the **interest rate** does not significantly impact coin demand, which contrasts with prior research (Mukhlis et al., 2020). This result may be attributed to coins' primary role in facilitating transactions rather than being driven by speculative or precautionary motives often associated with broad money or banknotes. In contrast, the estimates of **the dummy variable of holiday season** provide a negative and significant result at the 5% level across all models. During the Eid-al-Fitr holiday season, demand for coinage decreased. This conclusion contradicts the results of Lubis et al. (2019), who observed that currency demand will rise during holiday seasons such as Eid al Fitr and Christmas. As is customary during the holiday season, new banknotes are donated for charity causes, whilst coins get little attention. A cash-based society and non-economic elements may contribute to this phenomenon (Khiaonarong & Humphrey, 2023; Wisniewski et al., 2021). Our finding is corroborated by Skibińska-Fabrowska (2023) who assert that GDP and nominal interest rate influence cash demand, particularly high-denomination banknotes.

*4.2.3 Short-run Dynamis: Error Correction Model*

Table 4 shows the results for the short-run estimations. Across all three specifications, the estimated error correction terms (ECT) are statistically significant with negative signs. The negative sign of these variables signifies convergence towards long-run equilibrium, which supports the Bound test results, which suggested a cointegration relationship (Pesaran et al., 2001). The error correction term in all equations suggests that approximately 9.0 percent, 8.9 percent, and 6.8 percent of the deviations from the long-run equilibrium of the coin demand function, the error affects the short-term dynamics, is corrected to its long-run equilibrium in the subsequent period.

---

Insert Table 4

**Short-run Estimation Results**

---

The short-run ARDL estimates shown in Table 4 indicate that all macroeconomic variables proposed by the money demand theory were statistically significant in the short term, except the interest rate.

As expected, the income variable significantly and positively affects the coin money demand, particularly in current and previous periods, and the purchasing power of coins. The demand for coins was also substantially affected by the pandemic in the short and long periods. During the pandemic, individuals tend to minimize their use of cash, especially coins, because of the perceived risk of virus transmission. However, both in the long and short term, COVID-19 was found to be positively affecting the demand for coins, which corroborates Lalouette & Esselink (2018) and Honda & Uesugi (2022).

The absence of comprehensive research in economic literature regarding the coin usage renders it impossible to offer a definitive explanation for this phenomenon. Consequently, further investigation is advised, particularly to evaluate people's behaviour when using coins, which may differ from people's behaviour when using bank notes. This notion also applies to variables such as interest rate, highlighting that coins may serve just for transaction purposes rather than precautionary or speculative objectives. While income positively affects the demand for coins in the long term, it negatively influences coin demand in the short run. This condition can be perceived as an economic occurrence, but non-economic factors can impact this situation. For instance, this analysis does not account for the cultural context in which transactions occur and the distinct behaviours of merchants and consumers (Kotkowski, 2023).

### *4.3 Forecast of Coin Money Demand*

Figure 2 illustrates the results of the coin demand forecast generated by Model 1. The coin demand exhibits a consistent and upward trajectory for the subsequent five years, as depicted by a slope that deviates substantially from negative values. The upward trend indicates a consistent and growing demand for coins, which may be attributed to factors such as income growth, consumer preference, and other determinants impeding this trajectory in the short to medium term.

Insert Figure 2

**Forecast of Coin Money Demand**

As depicted in Figure 2, the observed increase in coin demand has substantial ramifications for individuals, policymakers, and economic agents. Notwithstanding the swift expansion of payment system innovations and the concurrent surge in non-cash payments within the economy, the demand for coins for transaction purposes is still and will continue to rise in the coming years. Thus, the coin remains essential to the economy and is not expected to become extinct anytime soon.

### *4.4 Diagnostic and Parameter Stability Tests*

It is important to ensure the robustness and reliability of estimates by conducting diagnostic tests. Our study includes four diagnostic examinations - tests for serial correlation, functional form, normality, and heteroscedasticity. Based on the outcomes presented in Table 5, it can be concluded that there are no heteroscedastic errors, serial correlation, model misspecification, or non-normality of residuals.

Insert Table 5

**Diagnostic Tests Result**

The empirical investigation of the stability of the estimated coefficients of the error correction model is also recommended by Erdiaw-Kwasie et al. (2023). We asses the stability of the money demand models employing Cumulative Sum (CUSUM) and Cumulative Sum of Squares (CUSUMSQ) tests.

---

Insert Figure 3

**(a) CUSUM test using the demand for coin (CMD) of Model 1, (b) CUSUM-of-squares test using the demand for coin (CMD) of Model 1**

---

Figures 3, 4, and 5 present the parameter consistency checks using the CUSUM and CUSUMSQ tests for models 1 to 3, respectively. The CUSUM and CUSUMSQ test results show that both the CUSUM and CUSUMQ recursive residuals for all models consistently fall within the critical boundaries. This observation provides empirical evidence supporting the stability of the money demand model and suggest the model has been properly specified.

---

Insert Figure 4

**(a) CUSUM test using the demand for coin (CMD) of Model 2, (b) CUSUM-of-squares test using the demand for coin (CMD) of Model 2**

---

---

Insert Figure 5

**(a) CUSUM test using the demand for coin (CMD) of Model 3, (b) CUSUM-of-squares test using the demand for coin (CMD) of Model 3**

---

## 5. Conclusion and Recommendation

The debate surrounding the necessity of coin money in the digital era is seldom openly discussed within academic circles despite its prevalence as a frequent subject of discussion in the practical economy. The present study seeks to explore the influence of payment innovations and the pandemic on the stability of coin demand in Indonesia. The scope of the study encompasses an expansion of the coin demand function, incorporating several measures of payment system innovation.

Our findings indicate a long-run relationship between coin demand and its determinants, namely, payment innovation, the pandemic, snd the coin value depreciation. The rapid expansion of electronic payment platforms had a significant effect on the coins demand, except for e-money. All control variables significantly influence coin demand, except for the interest rate, due to the coin's function as a medium of exchange rather than for speculative or precautionary purposes typically associated with banknotes or digital currency. Several results contrast with earlier money demand studies, raising questions about the underlying causes that require further investigation. Hence, it is recommended to incorporate supplementary features, such as non-economic variables, including people's behaviour, demography, and idiosyncratic components, into the model. Additionally, it is evidence that cash, particularly the Rupiah coin, remains and will persist as a legitimate mode of payment in Indonesia, despite long-standing predictions that cash usage in the country will inevitably decline or even vanish.

The findings of our study provide numerous significant implications. First, this article emphasizes that policymakers must understand the driving factors of coin demand in both economic and non-economic contexts to improve coin production-related issues and coin circulation policies. Second, an analysis of the denomination structure of Rupiah is crucial in addressing the problem of ineffective coin circulation in the economy, which is primarily attributable to their exceptionally low value compared to common products and services.

Finally, given the limitations inherent in econometric time series analysis, it would be advantageous to expand the research to encompass the behaviours of consumers, merchants, and commercial banks regarding the use of coins and alternative payment methods. Additionally, employing snowball research to explore the underlying causes of the issues would be beneficial.

**Acknowledgement**

The authors declare no conflict of interest. The views expressed in this article are those of the author(s) and do not necessarily represent the views or policies of Bank Indonesia.

Insert Appendix – Table A1

**Table A1. Variable description**

## Tables and Figures

### Figure 1 (a) Coin Money in Circulation and Policy Rate, (b) Non-Cash Payment Instruments

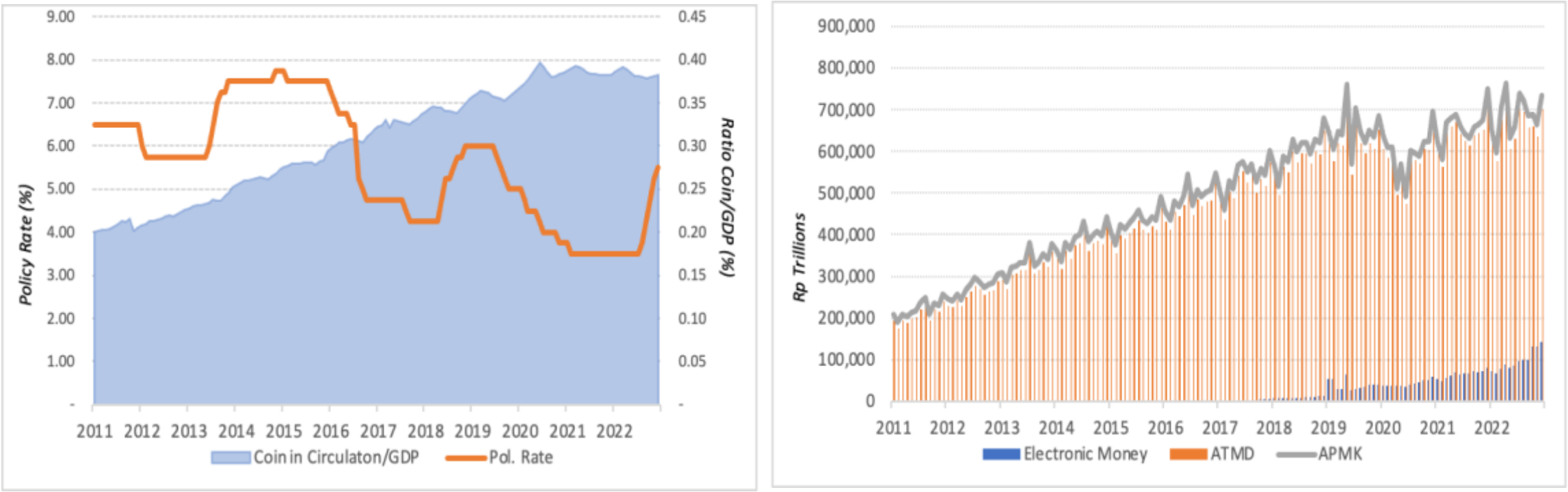

*Source: Bank Indonesia*

### Table 1. Descriptive Statistics

| NAME | STATISTICS | | | | UNIT ROOT TEST | | EXPECTED VALUE |
|---|---|---|---|---|---|---|---|
| | Mean | Min | Max | Stand. Dev. | Level | 1st Diff. | |
| LCMD | 29.591 | 28.878 | 30.066 | 0.368 | x | v | - |
| LGDP | 35.398 | 35.093 | 35.634 | 0.152 | x | v | + |
| IR | 5.986 | 2.790 | 8.630 | 1.524 | x | v | - |
| DUM_EID | 0.083 | 0.000 | 1.000 | 0.277 | v | v | + |
| DUM_C19 | 0.125 | 0.000 | 1.000 | 0.332 | x | v | - |
| VALCO | 0.176 | 0.107 | 0.255 | 0.038 | x | v | + |
| LAPMK | 33.766 | 32.872 | 34.270 | 0.365 | x | v | + |
| LATMD | 33.716 | 32.799 | 34.236 | 0.376 | x | v | + |
| LEM | 28.838 | 24.668 | 32.594 | 2.337 | x | v | - |

*Source: Bank Indonesia, Bloomberg, and author's calculation. All variables are in logarithmic form except for interest rate (IR), dummy variables, and value of coin (VALCO).*

**Table 2. ARDL Bound Test Results for Cointegration**

| No | Model specification | ARDL Order | F-statistic |
|---|---|---|---|
| 1 | F (LCMD, LGDP, IR, DUM_EID, DUM_C19, VALCO, LAPMK) | ARDL (1,3,0,0,3,2,0) | 12.203*** |
| 2 | F (LCMD, LGDP, IR, DUM_EID, DUM_C19, VALCO, LATMD) | ARDL (1,3,0,0,3,2,0) | 12.148*** |
| 3 | F (LCMD, LGDP, IR, DUM_EID, DUM_C19, VALCO, LEM) | ARDL (1,3,0,0,3,2,0) | 11.181*** |

*Source: Author's calculation*

**Table 3. Long-Run ARDL Estimation Results**

| | Model 1 | Model 2 | Model 3 |
|---|---|---|---|
| LOG (GDP (-1)) | 1.848*** | 1.857*** | 2.472*** |
| | [0.00] | [0.00] | [0.00] |
| IR | -0.008 | -0.008 | -0.008 |
| | [0.3536] | [0.347] | [0.507] |
| DUM_EID (-1) | -0.250** | -0.253** | -0.338** |
| | [0.016] | [0.017] | [0.03] |
| DUM_C19 | 0.108*** | 0.112*** | 0.136*** |
| | [0.001] | [0.001] | [0.0024] |
| log (VALCO)(-1) | 0.287** | 0.291** | 0.533** |
| | [0.014] | [0.0158 | [0.014] |
| log (ATMD) | 0.271** | | |
| | [0.0265] | | |
| log (APMK) | | 0.274** | |
| | | [0.031] | |
| log (EM) | | | 0.017 |
| | | | [0.581] |
| Constants | -44.367*** | -44.790*** | -57.315*** |
| | [0.00] | [0.00] | [0.00] |

*Source: Author's calculation. Notes: Figures in parentheses are p-value & *, **, *** indicate statistical significance at the 1, 5, and 10% level respectively.*

**Table 4.** Short-run Estimation Results

| | Model 1 | Model 2 | Model 3 |
|---|---|---|---|
| ΔLOG (CMD (-1)) | -0.090*** | -0.089*** | -0.068*** |
| | [0.001] | [0.001] | [0.007] |
| ΔLOG (GDP) | 0.255* | 0.254* | 0.271* |
| | [0.063] | [0.023] | [0.065] |
| ΔLOG (GDP (-1)) | 0.345** | 0.347** | 0.389** |
| | [0.021] | [0.014] | [0.01] |
| ΔLOG (GDP (-2)) | -0.397*** | -0.397*** | -0.425*** |
| | [0.003] | [0.001] | [0.002] |
| ΔIR | -0.001 | -0.001 | -0.001 |
| | [0.361] | [0.354] | [0.501] |
| ΔDUM_EID | -0.001 | -0.001 | -0.001 |
| | [0.598] | [0.501] | [0.71] |
| ΔDUM_EID (-1) | 0.015*** | 0.015*** | 0.015*** |
| | [0.00] | [0.00] | [0.00] |
| ΔDUM_EID (-2) | 0.011*** | 0.01*** | 0.01*** |
| | [0.00] | [0.00] | [0.00] |
| ΔDUM_C19 | 0.009*** | 0.009*** | 0.009** |
| | [0.008] | [0.007] | [0.019] |
| Δlog (VALCO) | 0.051* | 0.051** | 0.067*** |
| | [0.012] | [0.011] | [0.002] |
| Δlog (VALCO(-1)) | -0.043** | -0.043** | -0.037* |
| | [0.029] | [0.03] | [0.095] |
| Δlog (ATMD) | 0.024** | | |
| | [0.027] | | |
| Δlog (APMK) | | 0.024** | |
| | | [0.031] | |
| Δlog (UE)(-1) | | | 0.001 |
| | | | [0.555] |
| ECT (-1) | -0.09*** | -0.089*** | -0.068*** |
| | [0.00] | [0.00] | [0.00] |
| R-Squared | 0.425 | 0.424 | 0.404 |

Notes: Figures in parentheses are p-value & *, **, *** indicate statistical significance at the 1, 5 & 10% level respectively.

**Figure 2. Forecast of Coin Money Demand**

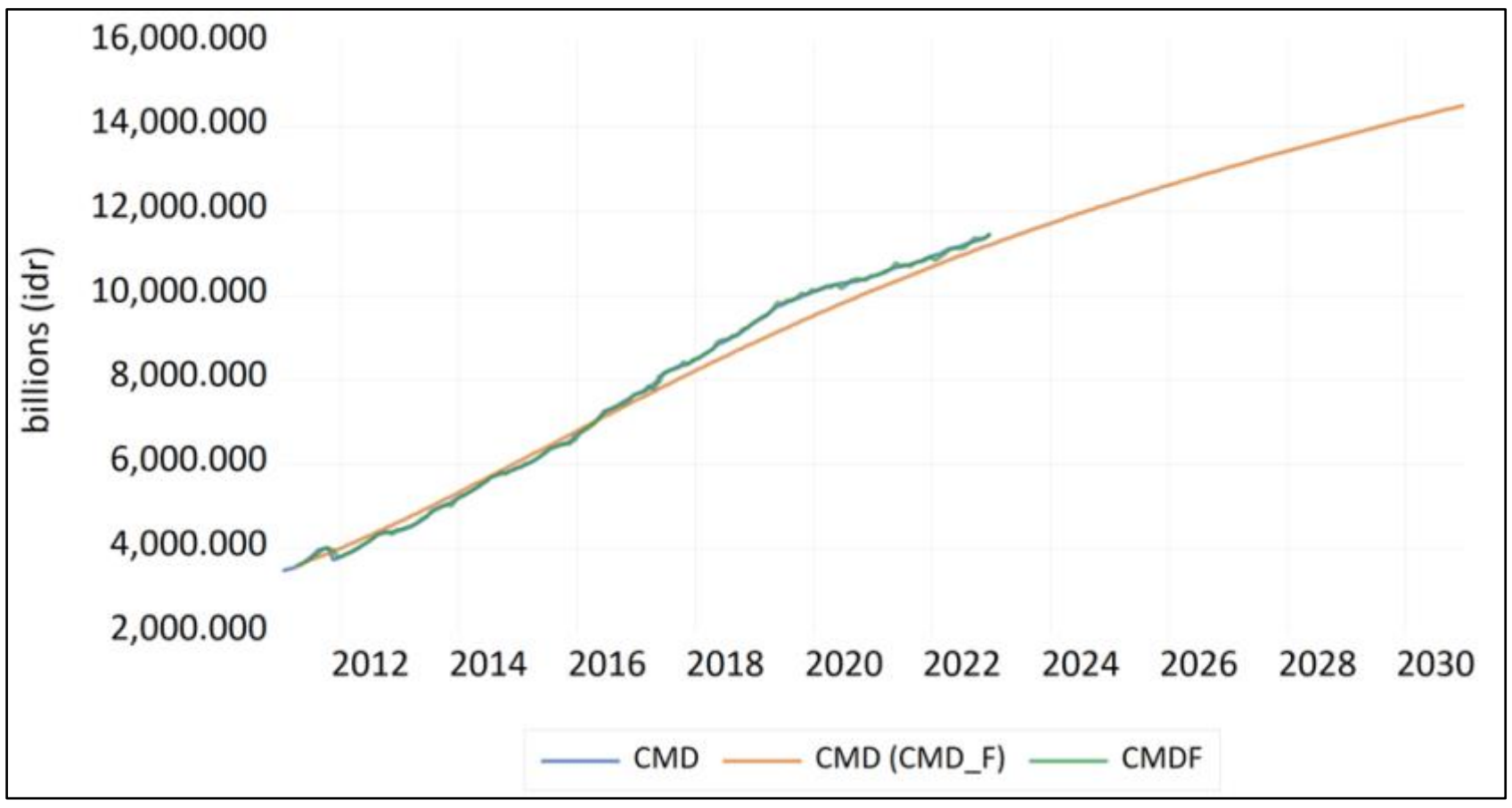

*Source: Author's calculation. Notes: CMD, CMD_F, and CMDF indicate actual CMD and CMD forecast using VAR and CMD prediction using ARDL, respectively.*

**Table 5. Diagnostic Tests Result**

| Diagnostic Statistic | Model 1 | Model 2 | Model 3 |
|---|---|---|---|
| R-squared | 0.425 | 0.519 | 0.403 |
| Adjusted R-square | 0.39 | 0.38 | 0.36 |
| F- Statistic (prob. Value) | 0.000*** | 0.000*** | 0.000*** |
| BG serial correlation LM test | 0.591 | 0.584 | 0.471 |
| Heteroskedasticty Test | 0.252 | 0.295 | 0.2752 |
| CUSUM | Stable | Stable | Stable |
| CUSUMSQ | Stable | Stable | Stable |
| Bound Test | 18.050*** | 19.227*** | 19.294*** |

*Source: Author's calculation*

**Figure 3 (a) CUSUM test using the demand for coin (CMD) of Model 1, (b) CUSUM-of-squares test using the demand for coin (CMD) of Model 1**

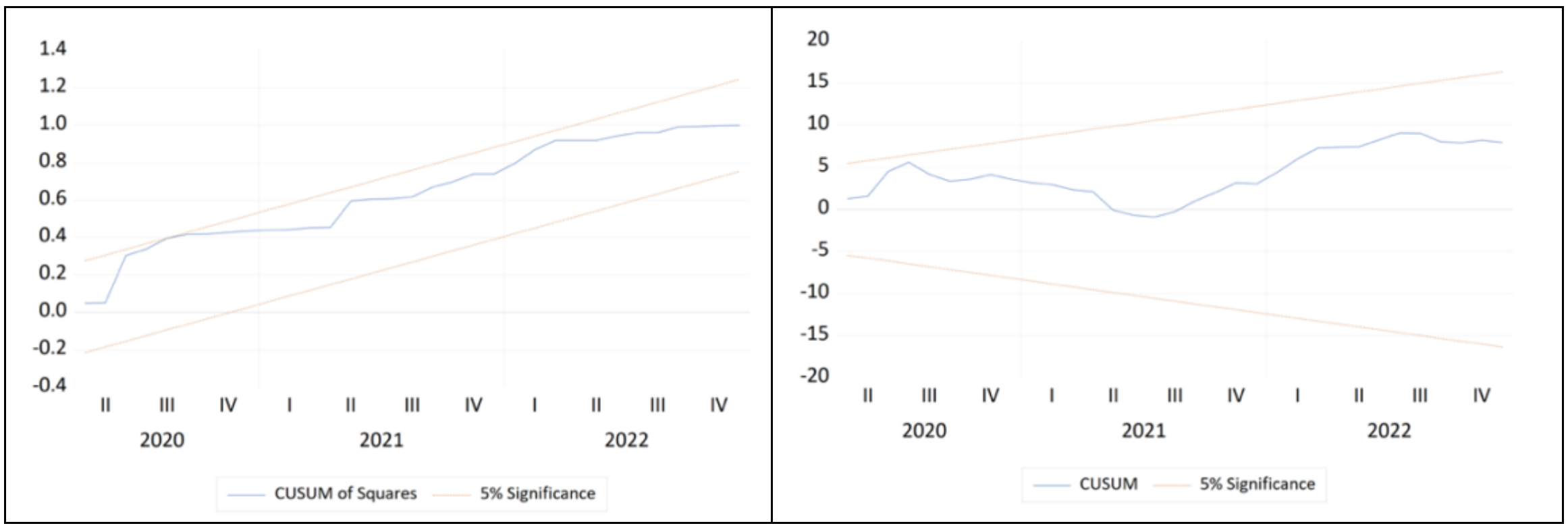

*Source: Author's calculation*

**Figure 4 (a) CUSUM test using the demand for coin (CMD) of Model 2, (b) CUSUM-of-squares test using the demand for coin (CMD) of Model 2**

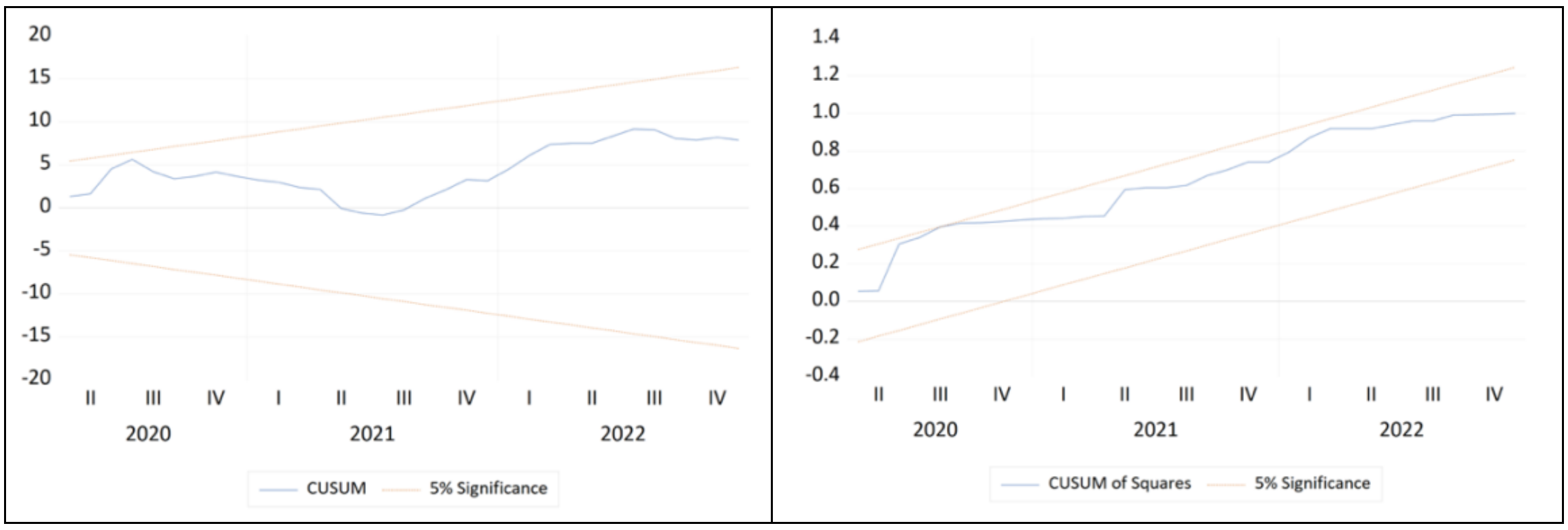

*Source: Author's calculation*

**Figure 4 (a) CUSUM test using the demand for coin (CMD) of Model 3, (b) CUSUM-of-squares test using the demand for coin (CMD) of Model 3**

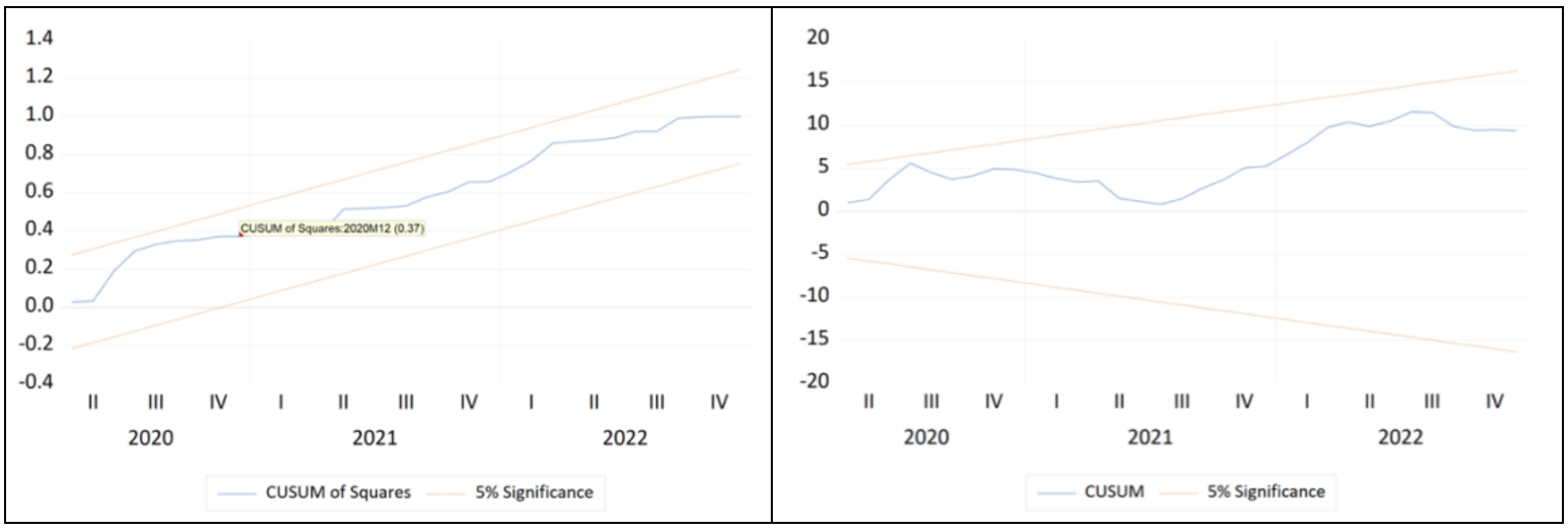

*Source: Author's calculation*

## Appendix

***Table A1. Variable description***

| No. | Label | Short Description | Abbreviation |
|---|---|---|---|
| 1 | Coin in circulation | The logarithm of Coin in Circulation (y to y) in Millions of Rp (LCMD) | CMD |
| 2 | Real GDP | The logarithm of gross domestic product (y to y) in Millions of Rp (LGDP) | GDP |
| 3 | Interest Rate | Policy Rate (IR) | IR |
| 4 | ATM/Debit Card | The logarithm of the value of ATM and debit card transactions, in Trillions of Rp (LATMD) | ATMD |
| 5 | Electronic Money | The logarithm of the value of electronic money (e-money) transactions, in Trillions of Rp (LEM) | EM |
| 6 | Payment System Innovation | The logarithm of the total value of the card and digital payment usage consists of ATM and debit cards, credit cards, and e-money in Trillions of Rp (LAPMK) | APMK |
| 7 | Value of the coin | Ratio of the highest Rupiah coin denominations compare to gold price | VALCO |
| 8 | Dummy of Eid Al-Fitr | A dummy to capture the Indonesian holiday of Eid al-Fitr is formed with indicator function = 0 for the normal period and the indicator function = 1 for Eid Al-Fitr. | DUM_EID |
| 9 | Dummy of COVID-19 | A dummy to capture the pandemic is formed with indicator function = 0 for periods 2011M1 to 202019M12 and the indicator function = 1 for periods 2020M1 and forward. | DUM_C19 |

*Source: Bank Indonesia, BPS, Bloomberg*

***Figure A2. Ratio of inflow to outflow of coins in Indonesia***

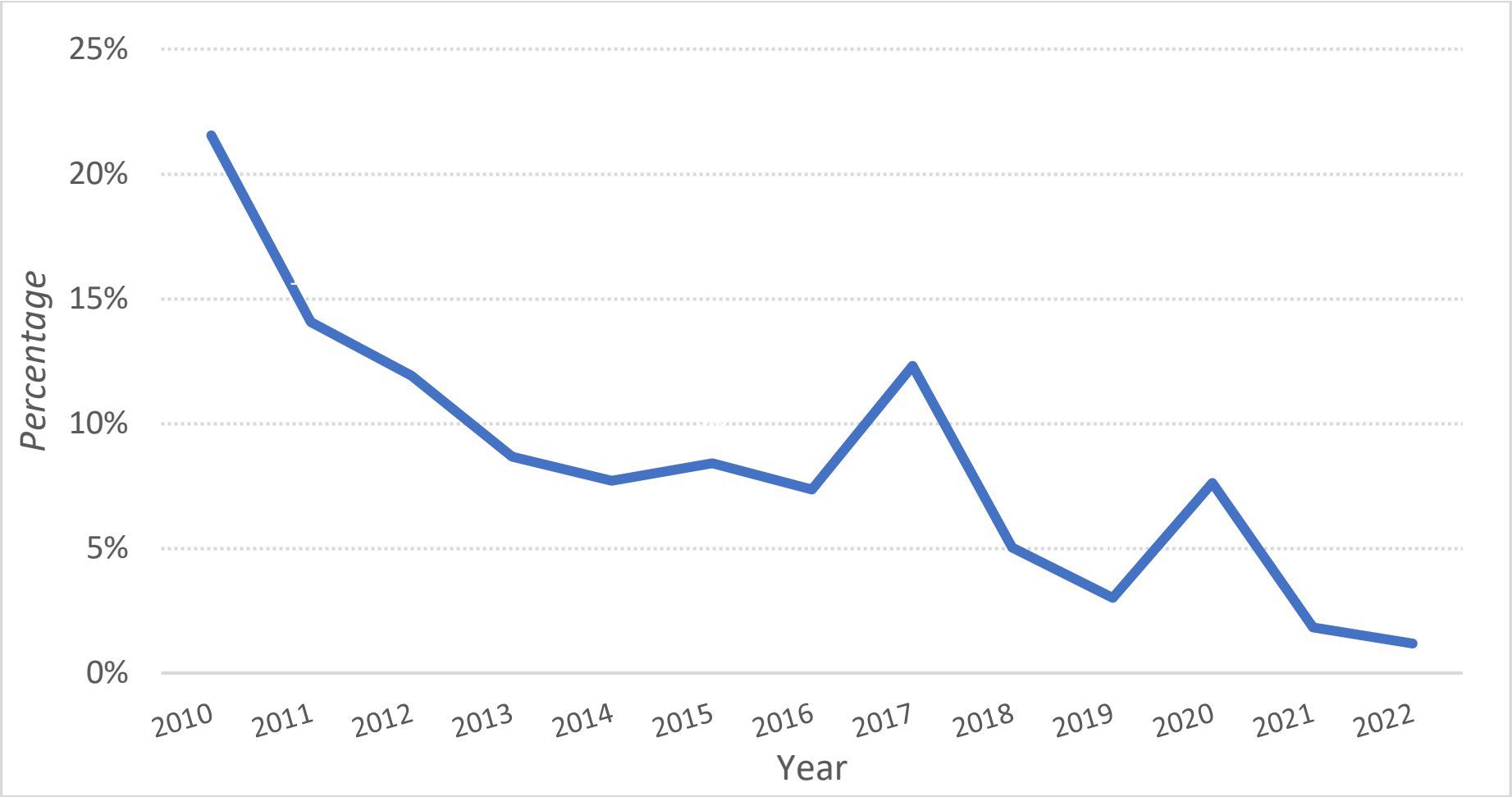

*Source: Bank Indonesia*